\documentclass[acmlarge]{acmart}

\usepackage{multicol}
\usepackage{alltt}

\usepackage{syntax}

\usepackage{booktabs}   
\usepackage[caption=false]{subcaption} 
                        
\usepackage{float}
\floatstyle{boxed} 
\restylefloat{figure}

\mathchardef\mhyphen="2D

\makeatletter\if@ACM@journal\makeatother

\startPage{1}
\else\makeatother

\startPage{1}
\fi

\setcopyright{none}             

\acmPrice{}

\makeatletter
\def\cf{%
  \begingroup
  \def\name{}%
  \cf@
}

\def\cf@{%
  \futurelet\ntoken\cf@@
}

\def\cf@@{%
  \ifcase 0%
    \ifx\ntoken\@sptoken 0\else
    \ifcat a\ntoken 1\else
    \ifcat 0\ntoken 2\fi
    \fi\fi
  \relax
    \expandafter\cf@end
  \or
    \expandafter\cf@add
  \else
    \expandafter\cf@checknum
  \fi
}

\def\cf@checknum#1{%
  \ifcase 0%
    \ifnum`#1>47
    \ifnum`#1<58 1\fi\fi
  \relax
    \def\next{\cf@end#1}%
  \else
    \def\next{\cf@add{#1}}%
  \fi
  \next
}

\def\cf@add#1{%
  \edef\name{\name#1}%
  \cf@
}

\def\cf@end{%
  \let\type\name
  \edef\name{\name:}%
  \def\cf@end{%
    \edef\@tempa{%
        \endgroup
        \type\noexpand~%
        \noexpand\ref{\name}%
    }%
    \@tempa
  }%
  \scantokens{\expandafter\let\csname cf@colon\endcsname=:}%
  \@ifnextchar\cf@colon%
    {\expandafter\cf@\@gobble}%
    {\cf@}%
}
\makeatother

\begin{document}

\title{dxo: A System for Relational Algebra and Differentiation}



\author{Julie S. Steele}
\affiliation{
  \institution{Georgetown Day School, USA}            
}
\email{jshermansteele@gmail.com}          

\author{William E. Byrd}
\affiliation{
  \institution{University of Alabama at Birmingham, USA}            
}
\email{webyrd@uab.edu}          


\begin{abstract}
    We present \emph{dxo}, a relational system for algebra and differentiation, written in miniKanren. 
    \emph{dxo} operates over math expressions, represented as s-expressions. \emph{dxo} supports addition, multiplication, exponentiation, variables (represented as tagged symbols), and natural numbers (represented as little-endian binary lists).
    We show the full code for \emph{dxo}, and describe in detail the four main relations that compose \emph{dxo}. 
    We present example problems \emph{dxo} can solve by combining the main relations.
    Our differentiation relation, {\tt do}, can differentiate polynomials, and by running backwards, can also integrate. Similarly, our simplification relation, {\tt simpo}, can simplify expressions that include addition, multiplication, exponentiation, variables, and natural numbers, and by running backwards, can complicate any expression in simplified form. Our evaluation relation, {\tt evalo}, takes the same types of expressions as {\tt simpo}, along with an environment associating variables with natural numbers. By evaluating the expression with respect to the environment, {\tt evalo} can produce a natural number; by running backwards, {\tt evalo} can generate expressions (or the associated environments) that evaluate to a given value. {\tt reordero} also takes the same types of expressions as {\tt simpo}, and relates reordered expressions. 
\end{abstract}



\maketitle

\section{Introduction}

Consider this calculus problem:

\begin{quote}
    Find two different polynomials, $f(x)$ and $g(x)$, and two different natural numbers $a$ and $b$, such that $f'(a)=b$, and $g'(b)=a$.
\end{quote}
Differentiating polynomials is an easy calculus problem, but the problem above is more complicated because of the relationships between the polynomials, their derivatives, and the two natural numbers. 
We invite the reader to pause, try to find solutions to this problem, and to think about how these types of problems might be solved more generally. 

We have developed a relational algebra system, 
\emph{dxo}, that uses relational programming to solve problems like the one above. We show the \verb|run| expression for solving this problem in Section~\ref{highlevel:sec}. \emph{dxo} is a collection of four main relations: \verb|simpo| for simplification, \verb|do| for differentiation, \verb|evalo| for evaluation, and \verb|reordero| for permuting arguments. 
Implementing \emph{dxo} relationally makes it flexible. For example, the relation \verb|do| can differentiate polynomials with respect to some variable. Since \verb|do| is a relation, it can also integrate polynomials. Also, the expression to be differentiated and its derivative can both contain fresh logic variables. The relations \verb|simpo|, \verb|evalo|, and \verb|reordero| similarly benefit from this flexibility. 
%

We assume the reader is familiar with core miniKanren~\cite{reasoned2,byrdthesis,Byrd:2006fk} (\verb|==|, \verb|fresh|, \verb|conde|, \verb|run|), extended with disequality (\verb|=/=|) and \verb|absento| constraints~\cite{Byrd:2012:MLU:2661103.2661105}.
Detailed explanations to the core miniKanren language can be found in ~\citet{reasoned2}, \citet{byrdthesis}, and \citet{Byrd:2006fk}.  
Descriptions of disequality and \verb|absento| constraints can be found in \citet{Byrd:2012:MLU:2661103.2661105} and \citet{byrd2017}.

Section~\ref{highlevel:sec} gives a high-level explanation of \emph{dxo}, its uses, and its four main relations. Section~\ref{codeexplanation:sec} explains in detail the main relations. Section~\ref{openproblems} discusses some open problems and possible future work. Section~\ref{relatedwork} discusses related work. We conclude the paper in Section~\ref{conclusion}. Appendix~\ref{codeappendix:sec} contains the full implementation of \emph{dxo}. 
%
%



%

\section{High-Level Overview}\label{highlevel:sec}

\emph{dxo} is composed of four main relations, \verb|simpo|, \verb|do|, \verb|evalo|, and \verb|reordero|, that when used in combination can solve interesting differentiation math problems. Here are the four relations and their uses:


\begin{description}
\item[{\ttfamily\fontseries{b}\selectfont{(simpo comp simp)}}]{relates \verb|comp| and \verb|simp|, where \verb|comp| can be any arithmetic expression and \verb|simp| is an equivalent, fully simplified one;}

\item[{\ttfamily\fontseries{b}\selectfont{(do x expr deriv)}}]{relates a polynomial expression \verb|expr| with its derivative \verb|deriv|, where the derivative is with respect to \verb|x|;}

\item[{\ttfamily\fontseries{b}\selectfont{(evalo env expr value)}}]{relates an expression \verb|expr| with its value \verb|value|, where each variable in \verb|expr| is associated with a natural number by the environment \verb|env|;}

\item[{\ttfamily\fontseries{b}\selectfont{(reordero e1 e2)}}]{relates two equivalent expressions, \verb|e1| and \verb|e2|, by changing the order of subexpressions in an addition or multiplication in any level of the other expression.}
\end{description}

Figure~\ref{simpogrammar:fig} contains the grammar for expressions accepted by \verb|simpo|, \verb|evalo|, and \verb|reordero|, and Figure~\ref{dogrammar:fig} contains the grammar for polynomial expressions accepted as the \verb|expr| for \verb|do|. \verb|deriv| is a subset. 
The implementation of \emph{dxo} uses the relational arithmetic system created by Oleg Kiselyov, which is presented in \citet{reasoned2} and \citet{declarativearithmetic}.

\begin{figure}[H]
\ 
\begin{grammar}

<dxo-expression> ::= \ 
\alt <numeral-or-variable> 
\alt `(+ ' <dxo-expression> $\ldots$ `)'
\alt `(* ' <dxo-expression> $\ldots$ `)'
\alt `(^ ' <dxo-expression> <dxo-expression> `)' 

<numeral-or-variable> :: = <tagged-numeral> | <tagged-variable>

<tagged-variable> ::= `(var ' <symbol> `)'

<tagged-numeral> ::= `(num ' <numeral> `)'

<numeral> ::= `()' | `(0 . ' <positive-numeral> `)' | `(1 . ' <numeral> `)' 

<positive-numeral> ::= `(0 . ' <positive-numeral> `)' | `(1 . ' <numeral> `)' 
\end{grammar}
\caption{Grammar for general \emph{dxo} expressions accepted by {\tt simpo}, {\tt evalo}, and {\tt reordero}.}
\label{simpogrammar:fig}
\end{figure}

\begin{figure}[H]
\ 
\begin{grammar}

<polynomial-expression> ::= \ 
\alt <numeral-or-variable> 
\alt `(+ ' <polynomial-expression> $\ldots$ `)'
\alt `(* ' <polynomial-expression> $\ldots$ `)'
\alt `(^ ' <numeral-or-variable> <tagged-numeral> `)'
\end{grammar}
\caption{Restricted grammar for polynomial expressions accepted by {\tt do}.}
\label{dogrammar:fig}
\end{figure}



Using the \emph{dxo} relations, we can solve the problem proposed in the introduction: find two different polynomials, $f(x)$ and $g(x)$, and two different natural numbers $a$ and $b$, such that $f'(a)=b$, and $g'(b)=a$. We relate \verb|f| and \verb|g| with their derivatives, \verb|fd| and \verb|gd|, using \verb|do|. Then we use \verb|evalo| to evaluate these derivatives at \verb|a| and \verb|b| respectively (we do this by making one environment where \verb|x| is \verb|a| and one where \verb|x| is \verb|b|), and set the evaluation to \verb|b| and \verb|a| respectively. Last, we make sure \verb|f| and \verb|g| are different but both simplified and \verb|a| and \verb|b| are different. 
\begin{alltt}
(run 20 (f g envb enva) 
  (fresh (b a gd fd)
    (=/= f g)
    (=/= b a)
    (== `((x . ,b)) envb)
    (== `((x . ,a)) enva)
    (do 'x f fd)
    (simpo f f)
    (do 'x g gd)
    (simpo g g)
    (evalo enva fd b)
    (evalo envb gd a)))
\(\Rightarrow\)
'(
  ...
  ((num _.0) (var x)                         ; f\( =c\textrm{ (where }c\textrm{ is any natural number),}\) g\( =x\)
   ((x)) ((x 1)))                            ; b\( =0\), a\(=1\)
  ...
  ((var x) (^ (var x) (num (0 0 1 1)))       ; f\( =x\), g\( =x\sp{12}\)
   ((x 1)) ((x 0 0 1 1)))                    ; b\( =1\), a\( =12\)
  ...
)
\end{alltt}
Of the 20 outputs produced by the run expression, many had $b=0$ so we only showed two. The first shown answer shows $$\frac{d}{dx}[c]=0, \textrm{where }c \textrm{ is any natural number}\textrm{ and } \frac{d}{dx}[x] =1, \textrm{where }x=5$$
The second shown answer shows 
 $$\frac{d}{dx}[x]=1, \textrm{where }x=12 \textrm{ and } \frac{d}{dx}[x\sp{12}] =12, \textrm{where }x=1$$
The concise \verb|run| expression solving this problem shows how \emph{dxo} benefits from the expressiveness of relational programming. 

We showed a combination of the four main \emph{dxo} relations in solving the problem in the introduction. We will shortly demonstrate another way to combine these core relations in the definition of \verb|anydo|, below.

Let's use \verb|do| to differentiate the polynomial $x\sp{3} + x\sp{0}$. Mathematically, $$\frac{d}{dx}[x\sp{3} + x\sp{0}]=(x\sp{2}*3)+0$$ The equivalent call to \verb|do| succeeds:
 \begin{alltt}
(do 'x
    '(+ (^ (var x) (num (1 1))) (^ (var x) (num ())))        ; \(x\sp{3}+x\sp{0}=\) expr
    '(+ (* (^ (var x) (num (0 1))) (num (1 1))) (num ())))   ; \((x\sp{2}*3)+0=\) deriv
       
\end{alltt}   
The derivative $(x\sp{2}*3)+0$ is equivalent to $1*x\sp{2}*3$, so we might expect the call
\begin{alltt}
(do 'x
    '(+ (^ (var x) (num (1 1))) (^ (var x) (num ())))      ; \(x\sp{3}+x\sp{0}=\) expr
    '(* (num (1)) (^ (var x) (num (0 1))) (num (1 1))))    ; \(1*x\sp{2}*3=\) deriv
       
\end{alltt}
to succeed. Unfortunately, this call fails because \verb|do| requires the derivative to be in canonical form, $(x\sp{2}*3)+0$ in this case. This means some mathematically correct expression and derivative pairs fail as arguments to \verb|do|.

We created \verb|anydo| to fix this problem. \verb|(anydo expr deriv x)|, like \verb|do|, relates an expression with its derivative with respect to \verb|x|, except \verb|anydo| generalizes this to simplified, complicated, or reordered forms of \verb|expr| and \verb|deriv|. This relaxes the restriction on \verb|deriv| being in canonical form, making running ``backward'' more convenient. Calling \verb|anydo| with the same arguments as above succeeds:
\begin{alltt}
(anydo '(+ (^ (var x) (num (1 1))) (^ (var x) (num ())))     ; \(x\sp{3}+x\sp{0}=\) expr
       '(* (num (1)) (^ (var x) (num (0 1))) (num (1 1)))    ; \(1*x\sp{2}*3=\) deriv
       'x)  
\end{alltt} 


%
\verb|anydo| is centered around a call to \verb|do| with arguments similar to \verb|expr| and \verb|deriv|, \verb|ecomp| and \verb|dcomp|. \verb|expr| and \verb|ecomp| are similar in that they simplify to the same value, \verb|esimp|, making them equivalent. \verb|anydo| does the same for \verb|deriv| and \verb|dcorder|, with the additional step of reordering \verb|dcorder| to be \verb|dcomp|. 
\begin{alltt}
(define anydo
  (lambda (expr deriv x)
    (fresh (esimp dsimp ecomp dcomp dcorder)
      (simpo expr esimp)
      (simpo ecomp esimp)
      (do x ecomp dcomp)
      (reordero dcomp dcorder)
      (simpo dcorder dsimp)
      (simpo deriv dsimp))))
\end{alltt}

\section{\emph{dxo} Implementation Walk-through}\label{codeexplanation:sec}

In this section we explain in detail the four main relations in \emph{dxo}.\footnote{We have released the \emph{dxo} code under an MIT licence at \url{https://github.com/JShermanSteele/dxo} .}
\subsection{simpo}

\verb|(simpo comp simp)| relates \verb|comp| and \verb|simp|, where \verb|comp| can be any arithmetic expression and \verb|simp| is an equivalent, fully simplified one. \emph{Simplified} means making all following simplifications: 
\begin{itemize}
\item $v+0=v$;
\item $v*0=0$;
\item $v*1=v$;
\item $v^0=1$ ($v \neq 0$);
\item $v^1=v$;
\item $0^v=0$ ($v \neq 0$);
\item and $1^v=1$;
\end{itemize}
\noindent
where $v$ is any expression.
For example, let's simplify $0^5 + (2*1)$:  
\begin{alltt}
(run* (simp) (simpo `(+ (^ (num ()) (num (1 0 1) ))             ; \(0\sp{5}+2*1=\) comp
                        (* (num (0 1)) (num (1)))) 
                    simp))
\(\Rightarrow\)
'((num (0 1)))                                                  ; \(2=\) simp
\end{alltt}

\verb|simpo| has base cases of \verb|(== comp simp)| for \verb|comp| and \verb|simp| being the same number or variable. The three non-base cases, addition, multiplication, and exponentiation, deeply recursively simplify sub-expressions by checking for those simplifyable cases. 



If \verb|comp| is ground, \verb|(simpo comp simp)| will either succeed exactly once or fail because there is at most one way to simplify any concrete expression, and \verb|simpo| has no overlapping cases when running "forwards". 
If \verb|comp| is fresh, then \verb|simpo| could succeed, but if it is an impossible relation, simpo can try longer and longer \verb|comp|s, never succeeding, and loop forever.


An example of these behaviors 
is that running \verb|(simpo comp simp)| with \verb|comp| as $1^1$ and \verb|simp| as a logic variable succeeds because $1^1$ simplified is $1$. Running with \verb|comp| as a logic variable and \verb|simp| as (the unsimplified) $1^1$ diverges, searching for a \verb|comp| forever.

\begin{alltt}
(run* (simp) (simpo '(^ (num (1)) (num (1))) simp))       ; \(1\sp{1}=\) comp
\(\Rightarrow\)
'((num (1)))                                                     

(run 1 (comp) (simpo comp '(^ (num (1)) (num (1)))))       ; \(1\sp{1}=\) simp
\end{alltt}

We can also construct a case with a ground \verb|comp| and partially ground \verb|simp| in which \verb|simpo| will fail. If \verb|comp| is $1^1$, which simplifies to $1$, and \verb|simp| is any addition expression, which may include fresh variables, \verb|simpo| will fail finitely. 
\begin{alltt}
(run* (q) (simpo `(^ (num  (1)) (num (1)))             ; \(1\sp{1}=\) comp
                 `(+ . ,q)))                           ; simp is some addition expression
\(\Rightarrow\)
'()
\end{alltt}


\subsection{do}
\verb|(do x expr deriv)| relates a polynomial expression \verb|expr| with its derivative \verb|deriv|, with respect to \verb|x|. For example, running  \verb|do| with \verb|expr| and \verb|deriv| fresh finds integral/derivative pairs:
\begin{alltt}
(run 24 (expr deriv) (do 'x expr deriv))

\(\Rightarrow\)

‘(...
  ((^ (var x) (num (0 1)))                          ; \(x\sp{2}=\) expr
   (* (^ (var x) (num (1))) (num (0 1))))           ; \(x\sp{1}*2=\) deriv 
  ...
  ((^ (var x) (num (1 _.0 . _.1)))                    ; \(x\sp{a} \textrm{where }a\textrm{ is odd}=\) expr
   (*                                                 ; \(x\sp{a-1}*a=\) deriv
     (^ (var x) (num (0 _.0 . _.1)))
     (num (1 _.0 . _.1))))
  ...
  ((* (^ (var x) (num ())) (^ (var x) (num ())))      ; \(x\sp{0}*x\sp{0}=\) expr
   (+                                                 ; \(0*x\sp{0}+x\sp{0}*0=\) deriv
     (* (num ()) (^ (var x) (num ())))
     (* (^ (var x) (num ())) (num ()))))
)
\end{alltt}
The best way to understand \verb|do| is as a case analysis on \verb|expr|, which is either a variable, a number, an exponentiation, an addition, or a multiplication.

Since the derivative of a sum is the sum of the derivatives of its parts, when \verb|expr| and \verb|deriv| are sums, the sub-expressions of \verb|expr| and \verb|deriv| are pair-wise related using \verb|do|. Since the sum can have any positive number of terms, a helper relation, \verb|map-do-o|, relates each pair in the sums.

 If \verb|expr| is a multiplication, \verb|do| must use the multiplication rule that %
 $$\frac{d}{dx}(ab)=\frac{da}{dx}*b + a*\frac{db}{dx},$$ and recur down the list of sub-expressions being multiplied. To improve the efficiency this process, we wrote a helper relation, \verb|multruleo|, that relates the list of sub-expressions being multiplied and the multiplication’s derivative. If the list has length greater than one, \verb|multruleo| separates the first term $e1$ from the rest, $e2 * e3 * \ldots$. Applying the multiplication rule to $e1 * e2 * e3 * \ldots$ yields $\frac{d}{dx}[e1]* e2 * e3 * \ldots + e1 * \frac{d}{dx}[e2 * e3 * \ldots]$, which is recursive with \verb|multruleo| because $\frac{d}{dx}[e2 * e3 * \ldots]$ is the related derivative argument to \verb|multruleo| with $e2 * e3 * \ldots$ as the first argument. This is \verb|conde| the clause in \verb|do| for multiplication. 
 
\begin{verbatim}
((fresh (l e)
   (== expr `(* . ,l))       
   (letrec ((multruleo
             (lambda (l dd)
               (fresh (e e* d d* a b)
                 (conde
                   [(== l `(,e))(do x e dd)]
                   [(== l `(,e . ,e*))
                    (== e* `(,a . ,b))
                    (== dd `(+ (* ,d . ,e*) (* ,e ,d*)))
                    (do x e d)
                    (multruleo e* d*)])))))
     (multruleo l deriv))))
\end{verbatim}

We could recur through the multiplication by recurring with every shorter multiplication as an argument to \verb|do|, but our approach is simpler because it does not exit \verb|multruleo| while reccuring through \verb|expr|’s multiplication. 

If \verb|expr| is an exponentiation, the second subexpression in the exponent must be a number by \verb|do|’s grammar, so $\frac{d}{dx}[x ^ n] = (n *( x ^ {n -1}))$ where $n$ is any number. There are three clauses for constants,  $\frac{d}{dx}[x^0] = 0$, $\frac{d}{dx}[n^m]=0$, and $\frac{d}{dx}[n]=0$ where $n$ and $m$ are any numbers so long as they both are not $0$. Finally, the derivative of just \verb|x| is one. 
	
Since \verb|do| orders \verb|deriv| a certain way (for example, $x\sp{2}*6$ instead of $6*x\sp{2}$), some integratable \verb|deriv|s will fail. This is why \verb|do| should be used with \verb|reordero|. For example,

\begin{alltt}
(run 1 (expr) 
  (do 'x expr '(* (^ (var x) (num (1))) (num (0 1))))) ; \(x\sp{1}*2=\) deriv 
\(\Rightarrow\)
'((^ (var x) (num (0 1))))                             ; \(x\sp{2}=\) expr 
\end{alltt}
produces an answer, but switching \verb|deriv|’s multiplication order diverges:
\begin{alltt}
(run 1 (expr) 
  (do 'x expr '(* (num (0 1)) (^ (var x) (num (1)))))) ; \(2*x\sp{1}=\) deriv
\end{alltt}

Similarly to \verb|simpo|, \verb|do| succeeds exactly once or fails running ``forwards'' when \verb|expr| and \verb|x| are ground. With \verb|expr| fresh, \verb|do| can succeed or can loop infinitely just like  \verb|simpo|.

%

\subsection{evalo}

\verb|evalo| evaluates, and is useful for solving equations.
\verb|(evalo env expr value)| relates an expression \verb|expr| with its value \verb|value|, where each variable in \verb|expr| is associated with a natural number by the environment \verb|env|. For example we can look for expressions that evaluate to $8$ in an environment that binds \verb|z| to $2$:
\begin{alltt}
(run 200 (expr) (evalo `((z . (0 1))) expr '(0 0 0 1)))      ; z\(=2, \)value\(=8\)
\(\Rightarrow\)
‘(...
  (* (var z) (num (0 0 1)))                                    ; \(z*4=\) expr
  ...
  (^ (num (0 1)) (num (1 1)))                                  ; \(2\sp{3}=\) expr
  ...
  (+ (var z) (num ()) (num (0 1 1)))                           ; \(z+0+6=\) expr
)    
\end{alltt}

\verb|evalo| deeply recurs through \verb|expr|, evaluating tagged little-endian binary lists into miniKanren numbers. For \verb|evalo|'s base cases when \verb|expr| is a variable or number, \verb|value| is the variable's value from \verb|env| or the miniKanren number respectively.
If \verb|expr| is an addition, multiplication, or an exponentiation, then \verb|evalo| relates the first term with its value \verb|evc|, the rest with its value \verb|evrest|, and adds, multiplies, or exponentiates \verb|evc| and \verb|evrest| to obtain \verb|value|.
 The \verb|evalo| code for addition does this:
\begin{verbatim}
((== `(+ ,c . ,rest) expr)
 (conde
   ((== '() rest) (evalo env c value))
   ((=/= '() rest)
   (evalo env c evc)
   (evalo env `(+ . ,rest) evrest)
   (pluso evc evrest value))))    
\end{verbatim}
This will recur through \verb|rest| and sum all the parts to relate \verb|expr| with \verb|value|. 

An interesting use of \verb|evalo| is to solve algebra problems by making \verb|env| fresh, for example looking for Pythagorean triples. So we set up $x^2+y^2=z^2$ and also a $z*z=\mathit{z\mhyphen squared}$ relation to make sure that \verb|z| is a natural number to find the classic Pythagorean triple!
\begin{alltt}
(run 1 (env)
  (fresh (xv yv zv z2v)
    (== `((x . ,xv) (y . ,yv) (z . ,zv) (z-squared . ,z2v)) env)
    (evalo env `(+ (^ (var x) (num (0 1))) (^ (var y) (num (0 1)))) z2v)
    (*o zv zv z2v)))    
\(\Rightarrow\)
'(((x) (y) (z) (z-squared)))                                           ; x\(=0, \)y\(=0, \)z\(=0\) 
\end{alltt}
Not what we wanted, alas.  Setting non-zero constraints, though, produces: 
\begin{alltt}
(run 1 (env)
  (fresh (xv yv zv z2v)
    (poso xv)
    (poso yv)
    (poso zv)
    (== `((x . ,xv) (y . ,yv) (z . ,zv) (z-squared . ,z2v)) env)
    (evalo env `(+ (^ (var x) (num (0 1))) (^ (var y) (num (0 1)))) z2v)
    (*o zv zv z2v)))   
\(\Rightarrow\)
'(((x 1 1)                                                             ; x\(=3\)
   (y 0 0 1)                                                           ; y\(=4\)
   (z 1 0 1)                                                           ; z\(=5\)
   (z-squared 1 0 0 1 1)))                                             ; z\(\sp{2}=25\)
\end{alltt}
which is a 3-4-5 right triangle. 

If either \verb|env| or \verb|expr| is fresh, \verb|evalo| can loop forever, trying more and more complicated \verb|env|s or \verb|expr|s. If both \verb|env| or \verb|expr| are ground, then \verb|evalo| will terminate since \verb|evalo| runs simply forwards in this case. 


\subsection{reordero}
\verb|(reordero e1 e2)| relates two equivalent expressions, \verb|e1| and \verb|e2|, by changing the order of subexpressions in an addition or multiplication in any level of the other expression. It is useful for taking integrals with \verb|do|. We can use \verb|reordero| to find all reorderings of an expression: 
\begin{alltt}
(run* (e2) (reordero `(+ (num (1)) (* (num (0 1)) (num (1 1)))) e2))   ; \(1+2*3=\) e1
\(\Rightarrow\)
'((+ (num (1)) (* (num (0 1)) (num (1 1))))                            ; \(1+2*3=\) e2
  (+ (* (num (0 1)) (num (1 1))) (num (1)))                            ; \(2*3+1=\) e2
  (+ (num (1)) (* (num (1 1)) (num (0 1))))                            ; \(1+3*2=\) e2
  (+ (* (num (1 1)) (num (0 1))) (num (1))))                           ; \(3*2+1=\) e2
  
\end{alltt}

\verb|(reordero e1 e2)| relates \verb|e1| and \verb|e2| by having the same outer operation (addition, multiplication, or exponentiation). For addition and multiplication the code is,
\begin{verbatim}
((fresh (o e1* e2*)
   (== `(,o . ,e1*) e1)
   (== `(,o . ,e2*) e2)
   (typeo o '+or*)
   (reorderitemso e1* e2*)))
\end{verbatim}
where \verb|e1*| and \verb|e2*| are permutations of each other. \verb|reorderitemso| checks that \verb|e1*| and \verb|e2*| have the same length, and then calls \verb|reorderinnero| on them. We created \verb|reorderitemso| to improve speed and divergence behavior of \verb|reordero| by requiring \verb|e1*| and \verb|e2*| have the same length before considering the relations in \verb|reorderinnero|. \verb|reorderinnero| relates permuted lists at any depth. To deeply reorder, \verb|reorderinnero| calls \verb|reordero| on the corresponding sub-expressions for \verb|e1*| and \verb|e2*|.


\begin{verbatim}
(define reorderinnero
  (lambda (e1* e2*)
    (fresh (c1 rc1 rest1 rest2)
      (conde
        ((== '() e1*) (== '() e2*))
        ((== `(,c1 . ,rest1) e1*)
         (removeo rc1 e2* rest2)
         (reorderinnero rest1 rest2)
         (reordero c1 rc1))))))
\end{verbatim} 
\verb|reordero| greatly reduces infinite loops because \verb|reorderitemso| checks that its arguments are the same length. This check keeps \verb|e1| and \verb|e2| the same structure and length at every depth, keeping the search finite. Checks like this would be useful to add other places in \emph{dxo} to reduce divergence. 

\section{Open Problems}\label{openproblems}
\verb|dxo| could be improved by expanding the grammars, improving the speed and termination, and using automatic differentiation. We would like to add expressions like $2\sp{x}$, $sin(x)$, and multiple variables. Currently, \emph{dxo} searches inefficiently, especially combinations of relations like \verb|anydo|, so we would like to speed these up. We would also like to make more calls terminate. We are interested in improving \verb|simpo|, possibly implementing Knuth-Bendix Completion\cite{knuth} relationally. We have done some preliminary work making a relation to replace \verb|do| that automatic differentiates forwards and backwards. 

\section{Related Work}\label{relatedwork}
Expresso \cite{expresso} is a computer algebra system written in Clojure using the miniKanren-inspired core.logic library. expresso's original intent was to be relational, but the author made it non-relational to include more advanced features.\cite{Maikcorrespondence} Like \emph{dxo}, it includes algebraic simplification, differentiation, and evaluation. Beyond \emph{dxo}, it includes rewriting in normal form and expressions like $sin$. 

The Reduce-Algebraic-Expressions system in Prolog \cite{Reduce-Algebraic-Expressions} is similar to \verb|simpo|, using certain simplification identities. simplifies expressions like $((x+x)/x)*(y+y-y)$ \(\Rightarrow\) $2*y$. It can make simplifications like $x+x$ \(\Rightarrow\) $2*x$ which \verb|simpo| cannot since \verb|simpo| currently only includes simplification rules involving $0$ and $1$. The Reduce-Algebraic-Expressions system is not relational.

\section{Conclusion}\label{conclusion}
\emph{dxo} applies relational programming to algebra and differentiation. It can differentiate, integrate, simplify, complicate, evaluate, create, and reorder. \emph{dxo} can concisely represent non-trivial math problems and find solutions. \emph{dxo} is a foundation for future exploration of relational programming in algebra.

\begin{acks}                            

We are grateful for the work of all the relational programmers whose work we have built upon. Our work would not have existed without the efforts of Dan Friedman and Oleg Kiselyov. In addition, we would like to thank Maik Schünemann for explaining his work. We would like to thank Brandon T. Willard and Alan T. Sherman for comments on a draft of this paper and sharing ideas with us. 
Finally, we thank the anonymous reviewers for their many helpful comments.

\end{acks}

\bibliography{paper}

\newpage
\appendix

\section{Full Implementation of \emph{dxo}}\label{codeappendix:sec}
\begin{alltt}
(require "faster-miniKanren/mk.rkt")
(require "faster-miniKanren/numbers.rkt")

;defines ^ as expt
(define ^ (lambda (a b) (expt a b)))

;defines ZERO and ONE
(define ZERO `(num ,(build-num 0)))
(define ONE `(num ,(build-num 1)))

;exponent: (^ a b)=c
(define expo
  (lambda (a b c)
    (fresh (bm1 rec)
      (conde
        ((== (build-num 0) b) (== (build-num 1) c))
        ((=/= (build-num 0) b)
         (pluso bm1 (build-num 1) b)
         (expo a bm1 rec)
         (*o a rec c))))))

;atom?
(define atom?
  (lambda (expr)
    (cond
      ((list? expr) #f)
      ((null? expr) #f)
      (else #t))))

;atom, null, or list 
(define typeo
  (lambda (expr answer)
    (fresh (a b)
      (conde
        ((== `(,a . ,b) expr) (== 'list answer) (=/= 'num a) (=/= 'var a))
        ((== `() expr) (== 'null answer))
        ((== `(num ,a) expr) (== 'atom answer))
        ((== `(var ,a) expr) (== 'atom answer))
        ((== '+ expr) (== '+or* answer))
        ((== '* expr) (== '+or* answer))))))
        
;miniKanren number
(define numo
  (lambda (n)
    (fresh (b rest)
      (conde
        ((== '() n))
        ((== `(,b . ,rest) n)
         (conde
           ((== 1 b))
           ((== 0 b)))
         (numo rest))))))
         
;empty env 
(define empty-env `())

;ext-env
(define ext-env
  (lambda (x v env)
    (cons `(,x . ,v) env)))

;lookupo
(define lookupo
  (lambda (x env v)
    (fresh (env* y w)
      (conde
        ((== `((,x . ,v) . ,env*) env))
        ((== `((,y . ,w) . ,env*) env) (=/= y x) (lookupo x env* v))))))
      
;unbuild-numinner to every element and if list, then to list
(define unbuild-numhelper
  (lambda (expr)
    (cond
      ((null? expr) '())
      ((list? (car expr)) (cons (unbuild-numinner (car expr)) (unbuild-numhelper (cdr expr))))
      (else (cons (car expr) (unbuild-numhelper (cdr expr)))))))

;calls unbuld-numinner for every answer in miniKanren 
(define unbuild-num
  (lambda (expr)
    (cond
      ((null? expr) '())
      (else (cons (unbuild-numinner (car expr)) (unbuild-num (cdr expr)))))))

;undoes build-num by calling unbinary 
(define unbuild-numinner
  (lambda (expr)
    (match expr
      [`() `()]
      [`(num ,b) `(num ,(unbinary b 1))]
      [`(var ,b) `(var ,b)]
      [`(+ ,e . ,e*) (unbuild-numhelper `(,e . ,e*))]
      [`(* ,e . ,e*) (unbuild-numhelper `(,e . ,e*))]
      [`(^ ,e . ,e*) (unbuild-numhelper `(,e . ,e*))])))

;helper for unbuild-numinner that goes from binary to base 10
(define unbinary
  (lambda (expr n)
    (cond
      ((null? expr) 0)
      ((atom? expr) expr)
      ((equal? (car expr) 1) (+ n (unbinary (cdr expr) (* 2 n))))
      ((equal? (car expr) 0) (unbinary (cdr expr) (* 2 n))))))


;l contains item 
(define membero
  (lambda (item l)
    (fresh (a rest)
      (conde
        ((== `(,item . ,rest) l))
        ((== `(,a . ,rest) l) (=/= item a) (membero item rest))))))

;l does not contain item
(define notmembero
  (lambda (item l)
    (fresh (a rest)
      (conde
        ((== '() l))
        ((== `(,a . ,rest) l)
         (=/= a item) 
         (notmembero item rest))))))
             
;removes item from l 
(define removeo
  (lambda (item contain removed)
    (fresh (rest rest2 c)
      (conde
        ((== `(,item . ,rest) contain)
         (== rest removed))
            
        ((== `(,c . ,rest) contain)
         (=/= item c)
         (== `(,c . ,rest2) removed)
         (removeo item rest rest2))))))
         

"SIMPLIFY";----------------------------------------------------------

(define simpo
  (lambda (comp simp)
    (fresh ()
      (conde
        ((fresh (n)
           (== `(num ,n) comp)
           (== comp simp)))
        ((fresh (v)
           (== `(var ,v) comp)
           (== comp simp)))
        
        ((fresh (e1 e2 s1 s2)
           (== `(^ ,e1 ,e2) comp)
           (conde
             ((== ONE s1) (== ONE simp))
             ((== ZERO s1) (=/= ZERO s2) (== ZERO simp))
             ((=/= ZERO s1) (== ZERO s2) (=/= ONE s1) (== ONE simp))
             ((=/= ZERO s1) (=/= ONE s1) (== ONE s2) (== s1 simp))
             ((== `(^ ,s1 ,s2) simp)
              (=/= ONE s1)
              (=/= ONE s2)
              (=/= ZERO s1)
              (=/= ZERO s2)))
           (simpo e1 s1)
           (simpo e2 s2)))

        ((fresh (e e* s temp t* n v)
           (== `(* ,e . ,e*) comp)
           (conde
             ((== '() e*) (simpo e simp))
             ((== ZERO s)(=/= '() e*)(== ZERO simp))
             ((== ONE s)(=/= '() e*) (simpo `(* . ,e*) simp))
             ((=/= ONE s)
              (=/= ZERO s)
              (=/= '() e*) 
              (conde
                ((== ZERO temp) (== ZERO simp))
                ((== ONE temp) (== s simp))
                ((== `(^ . ,n) temp) (== `(* ,s ,temp) simp))
                ((== `(+ . ,n) temp) (== `(* ,s ,temp) simp))
                ((== `(num ,n) temp) (=/= ZERO temp) 
                 (=/= ONE temp) (== `(* ,s ,temp) simp))
                ((== `(var ,v) temp) (== `(* ,s ,temp) simp))
                ((==`(* . ,t*) temp) (== `(* ,s . ,t*) simp)))
              (simpo `(* . ,e*) temp)))
           (simpo e s)))

        ((fresh (e e* s temp t* n v)
           (== `(+ ,e . ,e*) comp)
           (conde
             ((== '() e*) (simpo e simp))
             ((== ZERO s)(=/= '() e*)(simpo `(+ . ,e*) simp))
             ((=/= ZERO s)
              (=/= '() e*)
              (conde
                ((== ZERO temp) (== s simp))
                ((== `(^ . ,n) temp) (== `(+ ,s ,temp) simp))
                ((== `(* . ,n) temp) (== `(+ ,s ,temp) simp)) 
                ((== `(num ,n) temp) (=/= ZERO temp) (== `(+ ,s ,temp) simp))
                ((== `(var ,v) temp) (== `(+ ,s ,temp) simp))
                ((== `(+ . ,t*) temp) (== `(+ ,s . ,t*) simp)))
              (simpo `(+ . ,e*) temp)))
           (simpo e s)))))))


"DERIVATIVE";----------------------------------------------------------

;takes derivative        
(define do
  (lambda (x expr deriv)
    (fresh ()
      (symbolo x)
      (conde
        ((fresh (d* e* a b c d)
           (== expr `(+ . ,e*)) (== e* `(,a . ,b))
           (== deriv `(+ . ,d*)) (== d* `(,c . ,d))
           (samelengtho e* d*)
           (map-do-o x e* d*)))

        ((fresh ()
           (== expr `(^ (var ,x) (num ,(build-num 0))))
           (== deriv ZERO)))

        ((fresh (l e)
           (== expr `(* . ,l))       
           (letrec ((multruleo
                     (lambda (l dd)
                       (fresh (e e* d d* a b)
                         (conde
                           [(== l `(,e))
                            (do x e dd)]
                           [(== l `(,e . ,e*))
                            (== e* `(,a . ,b))
                            (== dd `(+ (* ,d . ,e*) (* ,e ,d*)))
                            (do x e d)
                            (multruleo e* d*)])))))
             (multruleo l deriv))))
        
        ((fresh (int intm1)
           (== expr  `(^ (var ,x) (num ,int)))
           (== deriv `(* (^ (var ,x) (num ,intm1)) (num ,int)))
           (minuso int (build-num 1) intm1)))

        ((fresh (int1 int2)
           (== expr `(^ (num ,int1) (num ,int2)))
           (== deriv ZERO)
           (conde
             ((poso int1))
             ((== ZERO int1)(poso int2)))))

        ((fresh ()
           (== expr `(var ,x))
           (== deriv ONE)))
                    
        ((fresh (int)
           (== expr `(num ,int))
           (== deriv ZERO)))))))

;maps do relation 
(define map-do-o
  (lambda (x expr* output)
    (fresh (e* e out out*)
      (conde 
        [(== expr* '()) (== output '())]
        [(== expr* `(,e . ,e*))
         (== output `(,out . ,out*))
         (do x e out)
         (map-do-o x e* out*)]))))

"EVALUATE";------------------------------------------------------------

;evaluater
(define evalo
  (lambda (env expr value)
    (fresh (m x c a b rest evc evrest eva evb)
      (conde
        ((== `(var ,x) expr) (lookupo x env value))
        ((== `(num ,m) expr) (numo m) (== m value))
        ((== `(+ ,c . ,rest) expr)
         (conde
           ((== '() rest) (evalo env c value))
           ((=/= '() rest)
            (evalo env c evc)
            (evalo env `(+ . ,rest) evrest)
            (pluso evc evrest value))))
        ((== `(* ,c . ,rest) expr)
         (conde
           ((== '() rest) (evalo env c value))
           ((=/= '() rest)
            (evalo env c evc)
            (evalo env `(* . ,rest) evrest)
            (*o evc evrest value))))
        ((== `(^ ,a ,b) expr)
         (evalo env a eva)
         (evalo env b evb)
         (expo eva evb value))))))

"REORDER";______________________________________________________________

;another option instead of using reordero is to always enter expressions in the same right order 
;reorders expression deeply, reordering any + and * expressions 
(define reordero
  (lambda (e1 e2)
    (fresh ()
      (conde
        ((== e1 e2) (typeo e1 'atom))
        ((fresh (o e1* e2*)
           (== `(,o . ,e1*) e1)
           (== `(,o . ,e2*) e2)
           (typeo o '+or*)
           (reorderitemso e1* e2*)))
        ((fresh (a1 b1 a2 b2)
           (== `(^ ,a1 ,b1) e1)
           (== `(^ ,a2 ,b2) e2)
           (reordero a1 a2)
           (reordero b1 b2)))))))


;permutes a list by calling reorderinnero, and calls reordero on the items in the list deeply
(define reorderitemso
  (lambda (e1* e2*)
    (fresh ()
      (samelengtho e1* e2*)
      (reorderinnero e1* e2*))))



;permutes and calls reordero on the items, helper for reorderitemso
(define reorderinnero
  (lambda (e1* e2*)
    (fresh (c1 rc1 rest1 rest2)
      (conde
        ((== '() e1*)(== '() e2*))
        ((== `(,c1 . ,rest1) e1*)
         (removeo rc1 e2* rest2)
         (reorderinnero rest1 rest2)
         (reordero c1 rc1))))))

"ANYDO";__________________________________________________________________________________

(define anydo
  (lambda (expr deriv x)
    (fresh (ecomp dcomp esimp dsimp dcorder)
      (project (expr deriv)
        (if (var? expr)
            (fresh ()
              (simpo deriv dsimp)
              (simpo dcorder dsimp)
              (reordero dcomp dcorder)
              (do x ecomp dcomp)
              (simpo ecomp esimp)
              (simpo expr esimp))
            
            (fresh ()
              (simpo expr esimp)
              (simpo ecomp esimp)
              (do x ecomp dcomp)
              (reordero dcomp dcorder)
              (simpo dcorder dsimp)
              (simpo deriv dsimp)))))))


(define doitallevalo
  (lambda (ieval inte deriv deval x env)
    (fresh (icomp dcomp isimp dsimp dcorder)
      (evalo env inte ieval)
      (evalo env deriv deval)
      (do x icomp dcomp)
      (reordero dcomp dcorder)
      (simpo deriv dsimp)
      (simpo dcorder dsimp)
      (simpo inte isimp)
      (simpo icomp isimp))))
\end{alltt}
\end{document}